\documentclass[twocolumn,showpacs,amsmath,amssymb,floatfix,pre]{revtex4}
\usepackage{graphicx}
\begin{document}
\title{Entropy of polydisperse chains: solution on the Husimi lattice} 
\author{Minos A. Neto}
\affiliation{Departamento de F\'{\i}ısica\\
Universidade Federal do Amazonas,
3000, Japiim, 69077-000, Manaus, AM\\
Brazil}
\author{J\"urgen F. Stilck}
\email{jstilck@if.uff.br}
\affiliation{Instituto de F\'{\i}sica\\
Universidade Federal Fluminense\\
Av. Litor\^anea s/n\\
24210-346 - Niter\'oi, RJ\\
Brazil}
\email{jstilck@if.uff.br}
\date{\today}

\begin{abstract}
We consider the entropy of polydisperse chains placed on a
lattice. In particular, we study a model for equilibrium
polymerization, where the
polydispersity is 
determined by two activities, for internal and endpoint monomers of a
chain. We solve the problem exactly on a Husimi lattice built with
squares and with arbitrary
coordination number, obtaining an expression for the entropy as a
function of the density of monomers and mean molecular weight of the
chains. We compare this entropy with the one for the monodisperse
case, and find that the excess of entropy due to polydispersity is
identical to the one obtained for the one-dimensional case. Finally,
we obtain a distribution of molecular weights with a rather complex
behavior, but which becomes exponential for very large
mean molecular weight of the chains, as required by scaling properties
which should apply in this limit.
\end{abstract}


\maketitle

\section{Introduction}
\label{intro}
The statistical mechanics of chains placed on lattices has a long
history, among the pioneering studies we may mention the study of
dimers on two-dimensional lattices as a model for adsorption of
diatomic molecules on a surface \cite{fr37}. In the simplest version
of the model, the dimers are chains with two monomers which occupy
edges between first neighbor sites of 
the lattice and only excluded volume interactions are considered, so
that the problem is athermal. In the sixties, the
entropy of dimers placed on two-dimensional lattices was calculated
exactly in the particular case in which the lattice is fully
occupied \cite{fkt61}. The extension of this calculation to
three-dimensional lattices or to situations where the lattice is not
fully occupied are still open problems. An interesting recent
contribution in the dimer problem is the extension of the solution to
the situation in 
which one site on the border of the lattice is not occupied
\cite{w06}, that is, the problem of dimers and one monomer. In another
generalization of the dimer problem a dimer is associated
to different energies if it is placed on different edges of an
anisotropic lattice \cite{nyb89}. These models lead to unusual phase
transitions and may be applied to the study of phase transitions in
some ferroelectric crystals (see \cite{sn74} for an example).

Another generalization is to consider linear chains with a
number of monomers (molecular weight) $M$. If the chains are totally
flexible, all allowed configurations will have the same energy and
again the model is athermal. One fundamental equation, in the
thermodynamical sense, for this gas of $M$-mers is the entropy per
site as a function of the fraction of lattice sites occupied by
monomers, which may be defined as:
\begin{equation}
s_M(\rho)=\lim_{V \to \infty} \frac{1}{V} \ln[\Gamma(N_p=\rho
V/M,M;V)],
\end{equation}
where $\Gamma(N_p,M;V)$ is the number of ways to place $N_p$ linear
chains with $M$ monomers each on the lattice with $V$ sites. In two
dimensions this problem has bee studied using series expansions
\cite{ncf89}, exact solutions on Bethe and Husimi lattices \cite{so90}
and transfer matrix solutions of the model defined on finite strips
followed by finite size scaling extrapolations to the two-dimensional
limit \cite{ds03}. 

Here we study a version of this model where the molecular weight of
the chains is not fixed. In other words, instead of the monodisperse
set of chains considered above, {\em polydisperse} chains, with
different numbers of monomers, are allowed. The particular model we
consider is defined in the grand-canonical ensemble and was used to
model equilibrium polymerization. It was applied by Wheeler and
co-workers to the polymerization transition of sulfur
\cite{wkp80}. The statistical weight of a configuration of chains
on the lattice is determined by the activities $z_i$ and $z_e$ for
internal and endpoint monomers of chains, respectively. The particular
case $z_i=0$ corresponds to the dimer model (no internal monomers
allowed), and in the limit $z_e \to 0$ only very long chains are
present. This second limit was considered in detail in
\cite{wkp80}, since a polymerization transition occurs in this
limit. Actually, the model may be mapped into the $n$-vector model of
magnetism in the limit $n \to 0$ and the magnetic field in the
magnetic model is associated to the activity of endpoint monomers
$z_e$ \cite{wkp80}.

For finite values of $z_e$, no phase transition is expected. Thermal 
generalizations of the model, where energies
are associated to different allowed configurations, were studied in
the literature with 
techniques similar to the one we will use here, with focus on phase
transitions which occur in these models, as well as on metastable states 
and glass transitions \cite{g03}. Also, a model where the sets of 
connected sites are not linear chains, but objects with loops, was also
considered before with emphasis on the percolation transition \cite{g01},
but in the particular case where this athermal model reduces to the one 
we study here no distinction is made between internal and endpoint 
monomers of the chains.

The solution of statistical mechanical models on hierarchical lattices
such as the Bethe lattice \cite{b82} is a useful method to estimate
the thermodynamic behavior of these models on real lattices
\cite{g95}, and the equilibrium polymerization model was solved  on
this lattice, both with and without dilution \cite{sw87}. However,
this work concentrated on the polymer limit of the model, where the
polymerization transition occurs. Recently, we studied the Bethe
lattice solution of the model in the general case \cite{ns08}. Besides
obtaining the entropy as a function of the density and the mean
molecular weight $\bar{M}$, we also studied the distribution of
molecular weights, which was found to be exponential, similar to the
distribution for the one-dimensional case \cite{snd06}. Here we
present a solution of the same model on a Husimi tree built with
squares. On the Bethe lattice, which corresponds to the core of a
Cayley tree, no closed loops are present, and this is one of the
relevant differences between this lattice and regular lattices. On the
Husimi tree we considered, small loops with four edges are present,
and therefore we may expect that the solution of models on this
lattice should be closer to the ones for hypercubic lattices than the
one on a Bethe lattice with the same coordination number, since
the elementary plaquette of both lattices will be the same. Another
motivation for the present calculation is that the exponential
molecular weight distribution which was always found on the Bethe
lattice solution may be an artificial result for this
lattice. Actually, since the critical exponents for such hierarchical
lattices are classical, an exponential distribution is expected in the
critical condition at the
polymer limit, since the classical value for the exponent $\gamma$ is
equal to unity \cite{s92}, but deviations from this distributions are
possible in the general case.

In section \ref{mods} we define the model more precisely and obtain
the entropy for a general Husimi lattice built with squares, as well
as the distribution of molecular weights. Final comments and
discussions may be found in section \ref{com}.

\section{Definition of the model and solution on the Husimi lattice}
\label{mods}
\begin{figure}[h]
\includegraphics[scale=0.6]{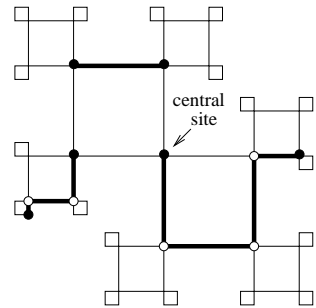}
\caption{A configuration of chains placed on a Husimi tree with
  branching parameter $\sigma=1$. Internal
  monomers are represented as white circles and endpoint monomers are
  black circles. The statistical weight of this configuration with 3
  chains is $z_i^5\,z_e^6$.} 
\label{f1}
\end{figure}

We consider a grand-canonical model of chains placed on a
lattice. Each chain is 
composed by two endpoint monomers, placed on the lattice sites, and
$0,1,2,\ldots$ internal monomers, corresponding to linear self-
and mutually avoiding walks on the lattice. The statistical weight of
a particular configuration of chains will be $z_e^{N_e}z_i^{N_i}$,
where $N_e$ and $N_i$ are the numbers of endpoint and internal
monomers in the configuration, respectively, while $z_e$ and $z_i$ are
the activities of the two types of monomers.
The partition function of this model on a lattice with $V$ sites may
be written as: 
\begin{equation}
\Xi(z_e,z_i;V)=\sum z_e^{N_e}z_i^{N_i}
\end{equation}
where the sum over all configurations
 of the chains on the lattice. 
In Fig. \ref{f1} a possible configuration of the chains is
shown. 
The density of endpoint monomers is
$\rho_e=\langle N_e \rangle/V$, the density of internal monomers is 
$\rho_i=\langle N_i \rangle /V$,
and the total density of monomers is $\rho=\rho_e+\rho_i$. The
densities may be obtained from the partition function as follows:
\begin{equation}
\rho_i=\frac{z_i}{V\Xi}\frac{\partial{\Xi}}{\partial z_i},
\label{ri}
\end{equation}
and
\begin{equation}
\rho_e=\frac{z_e}{V\Xi}\frac{\partial{\Xi}}{\partial z_e}.
\label{re}
\end{equation}

\begin{figure}[h]
\includegraphics[scale=0.6]{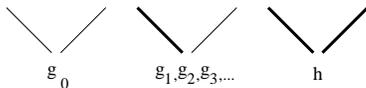}
\caption{The root configurations of the subtrees} 
\label{f2}
\end{figure}

Let us now consider the model defined on a Husimi tree (or cactus),
which is a Cayley tree built with squares. The branching parameter
will be equal to $\sigma$, so that the coordination number of the tree
is $q=2(\sigma+1)$. In Fig. \ref{f1} a tree with three generations of
squares is shown. The tree has a hierarchical
structure which allows many statistical mechanical models to be solved
in quite simple ways, similar to the ones used on Cayley trees and
Bethe lattices \cite{b82}. To solve the model on a Husimi tree, it is
convenient to consider rooted subtrees, with a square at the root
connected to $3\sigma$ other subtrees with one generation less. The
configuration of the two bonds incident on the root site of the
subtrees is fixed, and we define partial partition functions (ppf's)  
on these subtrees for fixed root configurations. The possible root
configurations are depicted in Fig. \ref{f2}, so that partial
partition function $g_0$ corresponds to a root configuration without
any bond coming from above, $g_i, i=1,2,3,\ldots$ are related to root
configurations with a single bond coming from above, connected to $i$
monomers. Finally, the partial partition function $h$ corresponds to
a root configuration with two incoming bonds. The root configurations with one
incoming bond have been split into the cases where the bond is already
attached to $i$ monomers since we later will be interested in
obtaining the distribution of sizes of the chains. It will be useful
to define the sum of all partial partition functions with a single
incoming bond:
\begin{equation}
k=\sum_{i=1}^\infty g_i.
\end{equation}

Now we obtain expressions for the partial partition functions of
subtrees with $n+1$ generations of squares in term of the ones of
subtrees with $n$ generations of squares. This may be done considering
the operation of attaching 3 groups of $\sigma$ smaller subtrees to a
new root square. As an example, the recursion relation for a subtree
without any bond incident on the root site will be:
\begin{eqnarray}
g_o^\prime &=& (g_0^\sigma+\sigma z_e g_0^{\sigma-1}k+
\frac{\sigma(\sigma-1)}{2}z_i g_0^{\sigma-2}k^2+\sigma z_i h)^3+ \nonumber \\
&&(z_eg_0^\sigma+\sigma z_i k g_0^{\sigma-1})^2\, 
[2(g_0^\sigma+\sigma z_e g_0^{\sigma-1}k+ \nonumber \\
&&\frac{\sigma(\sigma-1)}{2}z_i g_0^{\sigma-2}k^2+\sigma z_i h)+z_i g_0^\sigma].
\end{eqnarray}
The first term in the sum corresponds to a configuration of the root square 
without any polymer bond on its edges. Similar recursion relations may be 
obtained for the other ppf's, but since they are rather long they will not
be written down here. When these recursion relations are iterated, we obtain
the ppf's of a subtree with additional generations of sites. As expected, these
partition functions diverge in the thermodynamic limit. 

Let us now define the ratios of partial partition
functions:
\begin{subequations}
\begin{eqnarray}
G_i &=& \frac{g_i}{g_0}, \; i=1,2,\ldots \\
H &=& \frac{h}{g_0}, \\
K &=& \frac{k}{g_0}=\sum_{i=1}^\infty G_i.
\end{eqnarray}
\end{subequations}
As seen in the explicit recursion relation presented above, some linear 
combinations
of partial partition functions appear repeatedly, so it is convenient
to define them here:
\begin{eqnarray}
a &=& g_0^\sigma A=g_0^\sigma\left[1+\sigma z_e K+
  \frac{\sigma(\sigma-1)}{2}z_i K^2 \right. \nonumber \\
&&\left. +\sigma z_i H \right], 
\label{a}\\
b &=& g_0^{\sigma}B=g_0^{\sigma}(z_e+\sigma z_iK), 
\label{b}
\end{eqnarray}
The recursion relations for the ratios of partial partition functions are:
\begin{subequations}
\begin{eqnarray}
G_1^\prime &=& \frac{2z_e(A^2+B^2)}{D}, \\
G_2^\prime &=& \frac{2z_i[\sigma(A^2+B^2)G_1+z_eA]}{D}, \\
G_3^\prime &=& \frac{2z_i[\sigma(A^2+B^2)G_2+\sigma z_iAG_1+z_i z_e]}{D},
  \\
G_i^\prime &=& \frac{2\sigma z_i[(A^2+B^2)G_{i-1}+z_iAG_{i-2}+z_i^2
  G_{i-3}]}{D} \nonumber \\ 
&&(i \geq 4), \\
H^\prime &=& \frac{(A+2z_i)B^2}{D},
\label{rr1}
\end{eqnarray}
\label{rrall}
\end{subequations}
where $D=A^3+B^2(2A+z_i)$. Summing the recursion relations for the $G_i$
leads to:
\begin{eqnarray}
K^\prime&=&\sum_{i=1}^\infty G_i^\prime \nonumber \\
&=&\frac{2B(A^2+B^2+z_iA+z_i^2)}{D}.
\label{rr2}
\end{eqnarray}

In the thermodynamic limit, the behavior of the model will be
described by the fixed point values of the infinite set of
recursion relations Eqs. (\ref{rrall}). The linearity of these recursion
relations with respect to the ratios $G_i$ allows us to assure
that a fixed point of the pair of recursion relations Eqs. 
(\ref{rr1}) and (\ref{rr2}), $H^\prime=H,K^\prime=K$, is also
a fixed point of the whole set of recursion relations 
pair of recursion relations, so that we restrict ourselves to 
this reduced set of equations. Due to the non-linearity of these
equations, we were not able to find closed expressions of the fixed
point values $H(z_i,z_e;\sigma)$ and $K(z_i,z_e;\sigma)$ in general,
but their numerical determination is straightforward. Two particular
cases where the fixed point equations simplify are when only
dimers are allowed ($z_i=0$) and the polymer limit ($z_e \to 0$). In
these limits the recursion relations above reduce to results
in the literature \cite{so90}. Once the fixed point values of $K$ and
$H$ are found, it is easy to determine the remaining ratios
$G_1,G_2,\ldots$ using the equations above. The partition function of
the model on the Husimi tree may then be obtained considering the
operation of attaching $\sigma+1$ subtrees to the central
site: 
\begin{equation}
\Xi=g_0^{\sigma+1}\left[1+(\sigma+1)\left(z_eK+z_iH+
\frac{\sigma}{2}z_iK^2\right)\right].
\label{pf}
\end{equation}
We are not interested in the solution of the model on the whole Husimi
tree, which is expected to show a behavior which is quite different
from the one on regular lattices, and thus concentrate our attention
on the behavior at the core of the tree (Husimi lattice). Thus, we obtain
the densities of 
external and internal monomers at the central site, which are easily
found considering the contributions to the partition function in
Eq. (\ref{pf}) above. The results are:
\begin{equation}
\rho_e=\frac{z_eK}{\frac{1}{\sigma+1}+z_eK+z_iH+\frac{\sigma}{2}z_iK^2},
\label{rhoe}
\end{equation}
and
\begin{equation}
\rho_i=\frac{z_iH+\frac{\sigma}{2}z_iK^2}{\frac{1}{\sigma+1}+z_eK+z_iH+
  \frac{\sigma}{2}z_iK^2}.
\label{rhoi}
\end{equation}.

The entropy per site $s(\rho_e,\rho_i)$ is related to the activities
through the equations of state:
\begin{equation}
\ln z_e=-\left(\frac{\partial s}{\partial \rho_e}\right)_{\rho_i},
\label{se}
\end{equation}
and
\begin{equation}
\ln z_i=-\left(\frac{\partial s}{\partial \rho_i}\right)_{\rho_e}.
\label{si}
\end{equation}
The entropy may then be obtained inverting Eqs. (\ref{rhoe}) and
(\ref{rhoi}) 
to obtain $z_e(\rho_e,\rho_i)$ and $z_i(\rho_e,\rho_i)$ and then
performing the integration: 
\begin{equation}
s(\rho_e,\rho_i)=-\int_{0}^{\rho_e}\ln z_e(\rho,0) d\rho- \int_0^{\rho_i}\ln
z_i(\rho_e,\rho) d\rho.
\label{s}
\end {equation}
The first integral may be performed analytically, since in the dimer
limit the fixed point equations are exactly solvable. The resulting
expression is \cite{so90}:
\begin{eqnarray}
s(\rho_e,0) &=&
-(1-\rho_e)\ln(1-\rho_e)-\frac{\rho_e}{2}\ln\frac{\rho_e}{q}+
\nonumber \\
&&+\frac{1}{2}\left(\frac{q}{2}- 
\rho_e\right)\ln
\left(1-\frac{2\rho_e}{q}\right)+ \nonumber \\
&&\frac{1}{2}\left(\frac{q}{2}-
\rho_e\right)\ln(1+W)- \nonumber \\
&&-\frac{q}{8}\ln(1+2W-W^2),
\end{eqnarray}
where
\begin{equation}
W=\left(1-\frac{q}{2\rho_e}\right)+\left[\left(\frac{q}{2\rho_e}-1\right)^2
+1\right]^{1/2}
\end{equation}
and $q=2(\sigma+1)$ is the coordination number of the lattice. Now the
entropy may be found performing the second integration in
Eq. (\ref{s}) numerically.

We will, however, proceed along a different path, which leads to the
same results but with a simpler numerical calculation. We start with an
argument proposed by Gujrati \cite{g95} to obtain the bulk
grand-canonical free energy per site $\phi_b$, adapting it to a Husimi
lattice built with squares \cite{sg05}. This result may be obtained
assuming the free energy {\em per elementary square} to be a function of the
generation of the plaquette in the tree and that in the bulk it will
converge to a fixed value in the thermodynamic limit \cite{osb10}. A
simple derivation for the present case, where we suppose that the
center of the Husimi tree is a site, starts by associating a
generation number $m$ to each site of the tree, such that $m=0$ for
the central site, $m=1$ for the $3(\sigma+1)$ first neighbors of the
central site and so on, until the surface of the tree is reached for
$m=M$. The number of sites at the surface of the tree will be equal
to:
\begin{equation}
N_s(M)=3(\sigma+1)(3\sigma)^{M-1}.
\label{ns}
\end{equation}
The number of remaining sites in the tree, which are in the bulk, will
be:
\begin{equation}
N_b(M)=1+3(\sigma+1)\sum_{j=0}^{M-2}(3\sigma)^j=1+3(\sigma+1)
  \frac{(3\sigma)^{M-1}-1}{3\sigma-1}. 
\label{nb}
\end{equation}
Supposing that the free energy of the whole tree may be written as a 
sum of the contributions of its surface and the bulk, we have:
$\Phi(M)=-k_BT\ln \Xi(M)=N_b(M)\phi_b+N_s(M)\phi_s$, where $\phi_b$ and 
$\phi_s$ are the free bulk and surface free energies per site, respectively.
In the thermodynamic limit $M \to \infty$, we may then find that:
\begin{equation}
-\ln \Xi(M+1)+3\sigma \ln \Xi(M)=[N_b(M+1)-3\sigma N_b(M)]\varphi_b
\end{equation}
where $\varphi_b=\phi_b/(k_BT)$. The substitution of the expressions
(\ref{ns}) and (\ref{nb}) for the numbers of bulk and surface sites we
will find that:
\begin{equation}
\varphi_b=-\frac{1}{4}\ln\left(\frac{\Xi(M+1)}{\Xi(M)^{3\sigma}}\right).
\end{equation}
This expression is a generalization of the one obtained using a similar
argument by Semerianov and Gujrati for $\sigma=1$ (expression A7 in 
\cite{sg05}). 
If we now substitute the partition function (\ref{pf}) in the
expression above and observe that $g_0(M+1)/(g_0)^{3\sigma}$ is equal
to the fixed point value of the denominator of the recursion relations
(\ref{rr1}) $D$, so that
\begin{eqnarray}
\lefteqn{\varphi_b=}&&\nonumber\\
-\frac{1}{4}\ln \left[\frac{D^{\sigma+1}}{[1+(\sigma+1)(z_eK+z_iH+\sigma
    z_iK^2/2)]^{3\sigma-1}}\right],&&
\end{eqnarray}
where the thermodynamic limit is implicit since the fixed point values
of $D$, $K$, and $H$ are used. 

The entropy per site in the bulk of the tree is given by the state
equation:
\begin{equation}
s=-\left(\frac{\partial \phi_b}{\partial T}\right)_{\mu_e,\mu_i},
\end{equation}
which leads to the expression:
\begin{equation}
\frac{s}{k_B}=-\varphi_b-\rho_e\ln z_e-\rho_i\ln z_i.
\label{ent}
\end{equation}
Solving equations (\ref{rhoe}) and
(\ref{rhoi}) for the activities $z_e$ and $z_i$, we obtain the
equations: 
\begin{equation}
\rho_e=z_eK(\sigma+1)(1-\rho_e-\rho_i),
\label{eqze}
\end{equation}
and
\begin{equation}
\rho_i=z_i(\sigma+1)(H+\sigma K^2/2)(1-\rho_e-\rho_i).
\label{eqzi}
\end{equation}
To obtain the entropy as a function of the densities of internal
monomers and endpoint monomers, we numerically solve the set of four
equations (\ref{eqze}), (\ref{eqzi}), (\ref{rr1}), and (\ref{rr2})
(making $H^\prime=H$ and $K^\prime=K$ in the last two equations), so
that we obtain $z_e$, $z_i$, $H$, and $K$ as functions of the
densities, and then the entropy per site in the bulk using expression
(\ref{ent}). In general, this numerical procedure converges to the
fixed point and leads to accurate results, but there are problems when
the full lattice limit is approached, as expected, since the
activities diverge in this limit. Therefore, to compute the entropy 
in the full lattice limit we use a separate procedure we will describe 
below. We thus take the limit of diverging
fugacities keeping $z=z_e/z_i=\exp[(\mu_e-\mu_i)/(k_BT)]$ fixed. In
this limit, the variables defined in expressions (\ref{a}) and
(\ref{b}) diverge and it is convenient to define the new variables
\begin{subequations}
\begin{eqnarray}
A_\infty&=&\frac{A}{z_i}=\sigma zK+\frac{\sigma(\sigma-1)}{2}K^2
+\sigma H, \label{ainf}\\
B_\infty&=&\frac{B}{z_i}=z+\sigma K, \label{binf} \\
D_\infty&=&=\frac{D}{z_i^3}=A_\infty^3+B_\infty^3(2A_\infty+1), \label{dinf}
\end{eqnarray}
\label{inf}
\end{subequations}
and the recursion relations may be rewritten as
\begin{eqnarray}
H^\prime&=&\frac{B_\infty^2(A_\infty+2)}{D_\infty}, \label{rr1rc}
\\
K^\prime&=&\frac{2B_\infty(A_\infty^2+B_\infty^2+A_\infty+1)}{D_\infty}.
\label{rr2rc}
\end{eqnarray}
The density of endpoint monomers will be given by:
\begin{equation}
\rho_e=\frac{zK}{zK+H+\sigma K^2/2},
\label{rerc}
\end{equation}
and we notice that in the limit $z \to \infty$, which corresponds to
the lattice fully covered with dimers, we find that $\rho_e \to 1$,
while for $z \to 0$ we have $\rho_e \to 0$, as expected. The density
of internal monomers will be $\rho_i=1-\rho_e$, and the mean number of
monomers per chain is ${\bar M}=2/\rho_e$. Substitution of equation
(\ref{rerc}) in this last expression leads to:
\begin{equation}
H=\frac{K}{2}[z({\bar M}-2)-\sigma K].
\end{equation}
This last expression allows us to eliminate $H$ from the fixed point
equations associated to the recursion relations (\ref{rr1}) and
(\ref{rr2}), so that the fixed point equations reduce to:
\begin{eqnarray}
\frac{K}{2}[z({\bar M}-2)-\sigma K] D_\infty=
(A_\infty+2)B_\infty^2,\\
K D_\infty=2B_\infty[A_\infty(1+A_\infty)+B_\infty+1].
\end{eqnarray}
For fixed values of the mean molecular weight of the chains ${\bar
 M}$, these equations may numerically be solved for $K$ and $z$, using 
also the equations (\ref{inf}).  Once
the fixed point is found, we may obtain the entropy noticing that
expression (\ref{ent}) in the full lattice limit reduces to
$s/k_B=-\varphi_b-\rho_e \ln z - \ln z_i$, and substitution of the
bulk free energy in this limit allows us to write the entropy as:
\begin{equation}
\frac{s}{k_B}=\frac{1}{4}\ln\frac{D_\infty^{\sigma+1}}{[(\sigma+1){\bar
      M}zK/2]^{3\sigma-1}}-\frac{2}{{\bar M}} \ln z.
\end{equation}

\begin{figure}
\begin{center}
\includegraphics[scale=0.3]{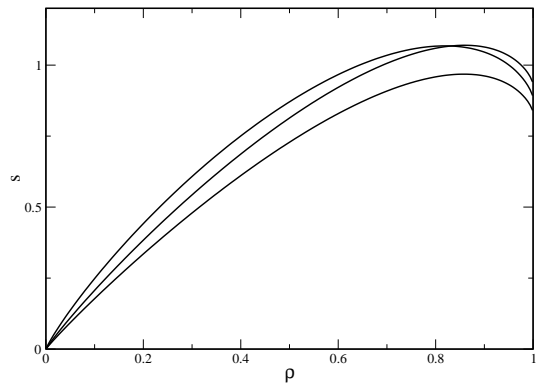}
\end{center}
\caption{Entropy as a function of the density of monomers
  $\rho=\rho_e+\rho_i$ for polydisperse chains on the Husimi
  lattice. In the region of small densities, the curves shown
  correspond to mean molecular weights ${\bar M}=3,6,$
  and $10$ in downward order. The results are for $\sigma=1$ ($q=4$).}  
\label{smr}
\end{figure}

\begin{figure}
\begin{center}
\includegraphics[scale=0.3]{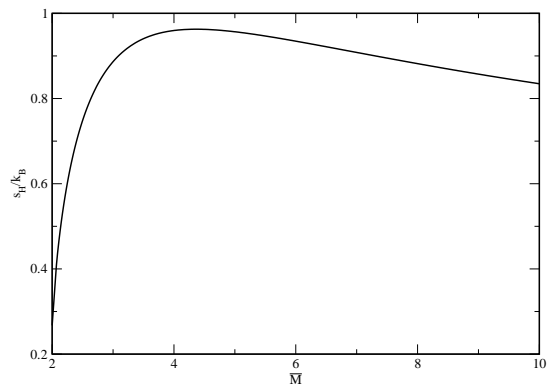}
\end{center}
\caption{Entropy as a function of the mean molecular weight at full
  coverage for a Husimi lattice with $\sigma=1$ ($q=4$).}
\label{shxm}
\end{figure}

On the Bethe lattice, it is possible to obtain an the entropy explicitly
using the expression for the bulk free energy, and we show this calculation
in some detail in the Appendix.

Results for the entropy as a function of the total density of sites,
for fixed values of the mean molecular weight ${\bar
  M}=2(\rho_e+\rho_i)/\rho_e$ are displayed in figure \ref{smr}. As
was already noticed in the results 
on the Bethe lattice, the entropy is a increasing function for small
values of the density and goes through a maximum at intermediate
densities. We notice that at low densities the entropy is a decreasing 
function of the mean molecular weight ${\bar M}$, but this does not
hold for larger densities. 

\begin{figure}
\begin{center}
\includegraphics[scale=0.3]{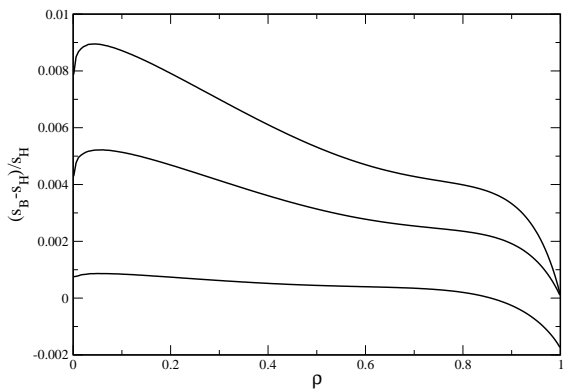}
\end{center}
\caption{Relative differences between the entropies of polydisperse
chains on Bethe ($s_B$) and Husimi ($s_H$) lattices as functions of 
the density for lattices with $q=4$. 
From the upper to the lower curve, the mean molecular weights are 
equal to 10, 6, and 3.}
\label{cent}
\end{figure}

\begin{figure}
\begin{center}
\includegraphics[scale=0.3]{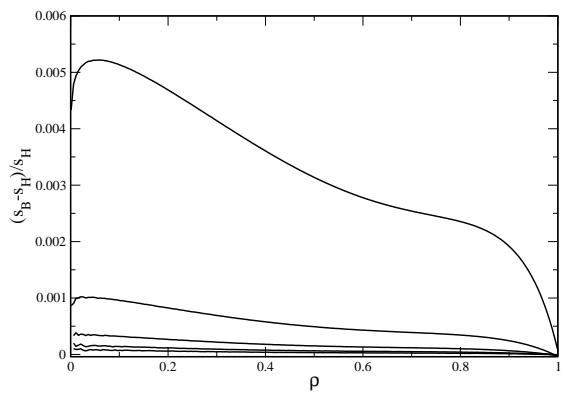}
\end{center}
\caption{Relative differences between the entropies of polydisperse
chains on Bethe ($s_B$) and Husimi ($s_H$) lattices as functions of the
density for ${\bar M}=6$. 
From the upper to the lower curve, the coordination numbers are equal to
4, 6, 8, 10, and 12. Numerical round-off errors are already visible at 
higher coordination numbers, particularly at low densities.}
\label{cents}
\end{figure}

It is interesting to compare the present 
results with the ones for the Bethe lattice \cite{ns08}. In general, as was 
also found in the monodisperse case \cite{so90}, both results are quite 
close. In figure \ref{cent} the relative differences between both entropies
are given as functions of the density for some values of ${\bar M}$, in general
these differences are in a range below 1\%. At low densities, the entropy on the
Bethe lattice is larger than the one on the Husimi lattice, but the opposite
may happen at higher densities. The difference between the results on both 
lattices reduces as the coordination number $q$ is increased, as expected,
since in the limit $q \to \infty$ both entropies are equal to the simple
mean field estimate, this is apparent in the plots displayed in figure
\ref{cents}. 

\begin{table}
\begin{tabular}{lllll}
\hline\hline
$M$&\textbf{Bethe (m)}&\textbf{Husimi
  (m)}&\textbf{Bethe (p)}&\textbf{Husimi
  (p)}\\ 
\hline\hline
$2$ &$\ \ 0,26162$     &$\ \ 0,26743$       &$\ \ 0,26162$     &$\ \ 0,26743$\\
$3$ &$\ \ 0,42284$     &$\ \ 0,41295$       &$\ \ 0,88493$     &$\ \ 0,88650$\\
$4$ &$\ \ 0,48166$     &$\ \ 0,48951$       &$\ \ 0,95904$     &$\ \ 0,95969$\\
$5$ &$\ \ 0,50669$     &$\ \ 0,50888$       &$\ \ 0,95656$     &$\ \ 0,95673$\\
$6$ &$\ \ 0,51349$     &$\ \ 0,51265$       &$\ \ 0,93473$     &$\ \ 0,93467$\\ 
$7$ &$\ \ 0,52217$     &$\ \ 0,52284$       &$\ \ 0,90836$     &$\ \ 0,90820$\\
$\infty$&$\ \ 0,4055$  &$\ \ 0,4090$        &$\ \ 0,4055$      &$\ \ 0,4090$\\
\hline\hline
\end{tabular}
\caption{Entropy at full coverage for lattices with $q=4$. Results for
monodisperse (m) and polydisperse (p) chains are shown. Data for the
monodisperse case are from reference \cite{so90} and the case of
polydisperse chains on the Bethe lattice is discussed in \cite{ns08}.}
\label{entrc}
\end{table}

In the table \ref{entrc} we present data for the entropy at
full coverage ($\rho=1$). As mentioned before for the general case,
the differences, both in
the monodisperse and the polydisperse systems, between the results of
the Bethe lattice and of the Husimi lattice solutions are quite
small, and there are indications that these values are
still rather far away from the results for regular lattices with the
same coordination number. For dimers ($M=2$), where there is no
polydispersity in our model, 
the exact value on the square lattice is known \cite{fkt61}
($s_2=G/\pi\approx 0.29156$, $G$ is 
Catalan's constant). In this case, the value on the Bethe lattice is
about 10\% below the exact value, while on the Husimi lattice the
difference reduces to roughly 8\%. The entropy for monodisperse
chains with $M>2$ calculated on Bethe and Husimi lattices is always
larger than the estimates obtained from transfer matrix calculations
for the square lattice \cite{ds03}, and the relative differences are
smaller than the ones for dimers. In the polymer limit $M \to \infty$,
the difference between the entropies for the poly- and monodisperse
cases vanishes again. 

We notice that the entropy at full coverage is not a
monotonic function of the mean molecular weight, showing a maximum
around ${\bar M} \approx 4.37$, as may be seen in figure
\ref{shxm}. As expected, on the Husimi lattice we find that the
entropy for polydisperse chains is always larger or equal to the
one for monodisperse chains with $M$ monomers for the same density
of monomers $\rho$ and finite ${\bar M}=M>2$, obtained in \cite{so90},
where equality holds only for vanishing density.

\begin{figure}
\begin{center}
\includegraphics[scale=0.3]{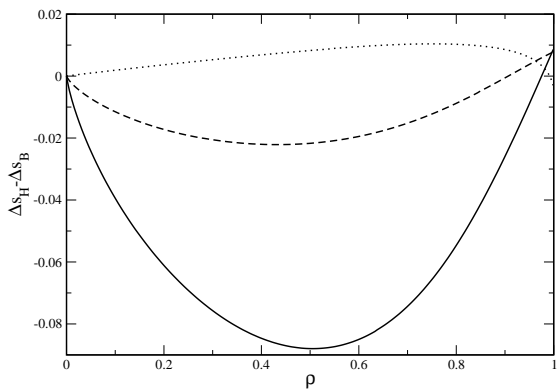}
\end{center}
\caption{Difference between the contribution of polydispersity
on the Bethe and the Husimi lattice solutions as a function of the
density $\rho$. The full curve is for $M=\bar{M}=3$, the dashed 
one for $M=\bar{M}=4$ 
and the dotted one for $M=\bar{M}=5$. All results are for 
$\sigma=1$ ($q=4$).}
\label{deltashb-m}
\end{figure}

\begin{figure}
\begin{center}
\includegraphics[scale=0.3]{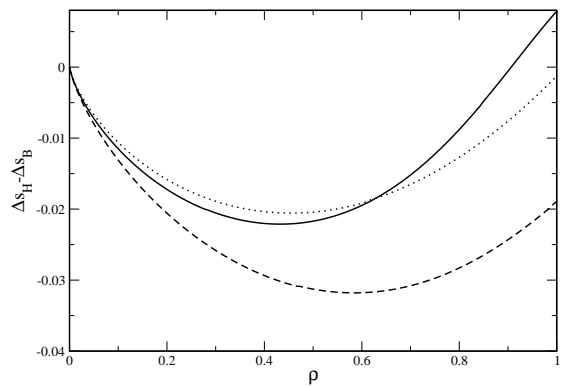}
\end{center}
\caption{Difference between the contribution of polydispersity
on the Bethe and the Husimi lattice solutions as a function of the
density $\rho$. The full curve is for $\sigma=1$, the dashed one for 
$\sigma=2$ and the dotted one for $\sigma=3$ ($q=6,4,8$, respectively).
All curves are for $M=\bar{M}=4$.}
\label{deltashb-s}
\end{figure}

On the Bethe lattice, it was 
possible to obtain the entropy for polydisperse chains analytically,
and it was found that the contribution of polydispersity to the
entropy, 
\begin{equation}
\Delta s_M(\rho)=s_{\bar{M}}(\rho)-s_M(\rho),
\end{equation}
with $M=\bar{M}$, is linear in $\rho$ and independent of $q$ \cite{ns08}:
$\Delta s_{M,B}=\rho[(M-1)\ln(M-1)-(M-2)\ln(M-2)]/M$. The contribution
to the entropy of polydispersity in the solution of the model on
the Husimi lattice ($\Delta s_{M,B}$ does not show such a simple 
behavior. In figure \ref{deltashb-m} we show the difference between
the results on the Bethe and on the Husimi lattices, for different values
of the molecular weight $M$. We notice that in general, as the 
molecular weight $M$ is increased, the Husimi lattice results are
closer to the ones found on the Bethe lattice. This may be understood
if we remember that since no loops are present on the Bethe lattice and
loops of four edges only may be found on the Husimi lattice we considered,
a reasonable effect of these closed paths should be expected for chains
of length close to four, but these effects should become smaller as
the chains grow. In opposition to what is found on the Bethe lattice,
the contribution of polydispersity to the entropy on the Husimi lattice
solution changes as the branching parameter $\sigma$ varies. This 
may be seen in figure \ref{deltashb-s}. In general, we find that the
Husimi lattice results become closer to the ones found on the Bethe 
lattice as $\sigma$, and therefore the coordination number $q$, grow. 
This is expected, since both solutions should become equal to the 
simple mean field solution in the limit $q \to \infty$. However, as 
is apparent in the curves in figure \ref{deltashb-s}, the convergence
is not monotonic.

\begin{figure}[h]
\vspace{0.5cm}
\includegraphics[scale=0.3]{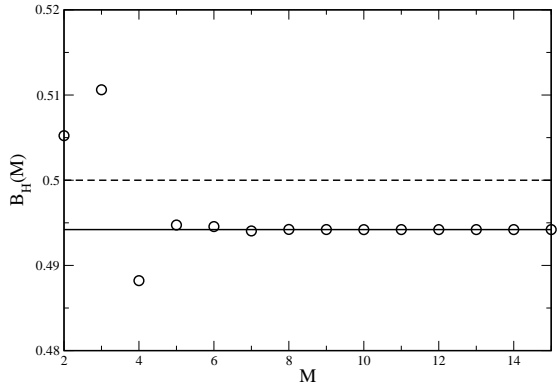}
\caption{Ratios of probabilities to find chains with successive
molecular weights $M$ on the Husimi lattice $B_H(M)$ as functions of the
molecular weight $M$ for ${\bar M}=3$. Data shown are for $\sigma=1$ 
and $\rho=0.3019045$. The dashed line corresponds to the result on the 
Bethe lattice, which for this case is $B_B=1/2$. The full line is the 
asymptotic value $B_H=\lim_{M \to \infty} B_H(M)$.}
\label{bhm}
\end{figure}

To calculate the probability $r_M$ to find a chain with $M$ monomers
among all chains, we notice that:
\begin{equation}
r_M=\frac{G_{M-1}}{K},
\end{equation}
and this probability can be numerically evaluated, since we may obtain
the functions $G$ and $K$, whose recursion relations are given in 
Eqs. (\ref{rr1}) and (\ref{rr2}), at the fixed point. 
We recall that on the Bethe lattice these ratios
are given by $B_B=r_{M+1}/r_M=({\bar M}-2)/({\bar M}-1)$, and therefore
are independent of $\rho$ and $q$ \cite{ns08}. 
As may be noted in Fig. \ref{bhm}, where the ratios $B_H(M)=r_{M+1}/r_M$ are
plotted as functions of $M$, for small values of $M$ compared to
4 the distribution of molecular weights is not exponential, but
as $M \gg 4$ an exponential behavior is apparent. The value 4
corresponds to the size of the only possible closed path on the lattice we
considered. Again, we notice that the deviations from the exponential
behavior are larger for smaller values of $q$, as $q$ increases the 
results approach an exponential behavior for all values of $M$, as found 
on the Bethe lattice. The asymptotic value of the ratio 
for large values
of $M$ may be found by assuming an exponential decay of the
probabilities for the fixed point values of the ratios of partial
partition functions $G_i$ for $i>3$ in Eqs. (\ref{rr1}), which leads to
the following equation for the limiting ratio $B_H=\lim_{M \to
\infty} r_{M+1}/r_M$:
\begin{equation}
B_H^3-\frac{2\sigma z_i}{D}[(A^2+B)\,B_H^2+z_iA\,B_H+z_i^2]=0,
\label{betaeq}
\end{equation}
and the horizontal line in Fig. \ref{bhm} was obtained solving this
equation. 

\begin{figure}[h]
\vspace{0.5cm}
\includegraphics[scale=0.3]{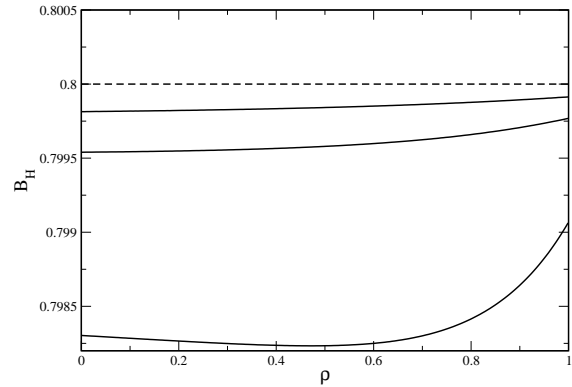}
\caption{Asymptotic coefficient $B_H$ as a function of the density $\rho$,
calculated for ${\bar M}=6$. The lowest curve corresponds to $\sigma=1$
$(q=4)$ and the following ones in upward order to $\sigma=2,\;3$ $(q=6,\;8)$.
The dashed line corresponds to coefficient for the Bethe lattice $B_B=4/5$.}
\label{bh}
\end{figure}

The asymptotic ratio of the solution on the Husimi lattice, $B_H$, in general 
will be a function of $\bar{M}$, $\sigma$ and $\rho$. In figure \ref{bh} 
this is apparent, and again we notice that as the coordination number
of the lattice become larger, the result on the Husimi lattice approaches 
the one found on the Bethe lattice. This may be seen analytically in the 
limit of vanishing
activities. Expanding the recursion relations up to the lowest nonzero
order of $z_e$ and $z_i$, we get:
\begin{subequations}
\begin{eqnarray}
H^\prime&=&z_e^2+2\sigma z_ez_iK+(\sigma z_i K)^2, \\
K^\prime&=&2z_e+2\sigma z_i K.
\end{eqnarray}
\end{subequations}
It is then easy to find the fixed point values up to lowest non-vanishing
order $K^*=2z_e$ and $H^*=z_e^2$, and the corresponding values for the
mean molecular weight $\bar{M}=2+(1+2\sigma)z_i$ and density of
monomers $\rho=2(\sigma+1)z_e^2$. The approximate solution of the
equation (\ref{betaeq}) for the decay exponent in the limit of small
activities is $\beta_H=2\sigma z_i$, which may be combined with the
expression for $\bar{M}$ above leading to:
\begin{equation}
\beta_H=\frac{2\sigma}{1+2\sigma}(\bar{M}-2),
\end{equation}
where the dependence of the exponent with $\sigma$ in the limit
considered is clear. In general, the
exponent will also be a function of $\rho$, but, as may be seen above,
in the limit we consider here this contribution is of higher
order. The same expansion may be done for the Bethe lattice solution
\cite{ns08}, and leads to the result $\beta_B=\sigma^\prime z_i$,
where $\sigma^\prime$ is the branching parameter of this lattice. If we
compare Bethe and Husimi lattices with the same coordination numbers,
we have $\sigma^\prime=2\sigma+1$, and as expected $\beta_H \to
\beta_B$ as $\sigma \to \infty$. 

\section{Final comments and discussions}
\label{com}
We have studied an athermal model of flexible chains with excluded volume
interactions only on a Husimi lattice built with squares, with an arbitrary
value for the branching parameter $\sigma$, so that the coordination 
number of the lattice is $q=2(\sigma+1)$. The chains are linear and composed 
by a set of monomers, so that consecutive monomers occupy first neighbor 
sites of the lattice. They are polydisperse, and 
the distribution of sizes is determined in an annealed way by two parameters:
the activity $z_e$ of an endpoint monomer (linked to one other monomer
of the chain only) and $z_i$ of an internal monomer, which is linked to 
two other monomer of its chain. 

The entropy as a function of the fraction of sites occupied by monomers and
the mean number of monomers in each chain (mean molecular weight of the 
chains) is a fundamental equation of the system in the thermodynamic sense, 
and we have obtained this function in general on the Husimi lattice, although, 
at variance to what was done in the solution of this problem on the Bethe
lattice \cite{ns08}, we were not able to derive a closed form for it. We found
that the differences between results on Bethe and Husimi lattices with the
same coordination numbers s rather small, below 1\%, as was also the case 
for monodisperse chains \cite{so90}. It is also possible that the results
on the Husimi lattice for $q=4$ are still not very close to the ones on the
square lattice, although, to our knowledge, no results for the entropy
of this model on regular lattices are available in the literature. For
the monodisperse rather precise estimates were obtained using the transfer
matrix approach \cite{ds03}, and we are presently extending these results 
for the polydisperse model on the square lattice. It is expected that the 
results on the Husimi lattice will be closer to the ones on regular lattices
at higher dimensions, but again there are no estimates of the entropy
of the model available in the literature for three-dimensional lattices,
for example. 

Another point which should be further investigated is the distribution of
molecular weights of the chains. The simple exponential distribution which
was found on the Bethe lattice \cite{ns08} is no longer valid on the Husimi 
lattice, so we expect that on regular lattices a different size distribution
will be found too. It may be that this question could be addressed on the
square lattice with transfer matrix techniques, we are now investigating this 
possibility.

\section*{Acknowledgments}
We acknowledge critical readings of the manuscript by Dr. Tiago J. Oliveira 
and Dr. Wellington G. Dantas. MAN acknowledges funding by FAPEAM and a 
doctoral grant from the Brazilian agency CNPq and JFS thanks the same 
agency for partial financial assistance.

\appendix
\section{Free energy of the model on Bethe lattice}
On a Bethe lattice with arbitrary coordination number $q$, the entropy
of chains whose polydispersity is determined by different activities
for internal and endpoint monomers may be calculated exactly. Here we
show that this result, which was originally obtained integrating the
expressions for monomer densities, may also be found more directly by
derivation of the bulk free energy calculated using Gujrati's
prescription. Recalling the discussion of the problem in \cite{ns08},
we define a partial partition functions $g_0$ and $g_1$
for subtrees with and without a polymer bond on the root edge,
respectively. The recursion relation for the ratio $R=g_1/g_0$ of
these ppf's is given by expression (13) in this paper:
\begin{equation}
 R^\prime=\frac{z_e+\sigma z_iR}{1+\sigma z_eR+
\frac{\sigma(\sigma-1)}{2} z_iR^2},
\end{equation}
and equation (14) for the partition function of the model on the
Cayley tree may be written as:
\begin{eqnarray}
\Xi&=&g_0^q+qz_eg_0^{q-1}g_1+\frac{q(q-1)}{2}z_ig_0^{q-2}g_1^2
\nonumber \\
&=&g_0^q\left[ 1+qz_eR+\frac{q(q-1)}{2}z_iR^2\right].
\label{gcpfb}
\end{eqnarray}
The free energy of the model on the Bethe lattice corresponds to free
energy on the bulk of the Cayley tree, and denoting this free energy
per site of the tree by $\phi_b$ we obtain, using Gujrati's ansatz,
the result \cite{oss09}:
\begin{equation}
\varphi_b=\frac{\phi_b}{k_bT}=-\frac{1}{2}\ln \frac{\Xi_{m+1}}
{\Xi_m^{q-1}},
\label{phibb}
\end{equation}
where $m$ denotes the number of generations in the tree and we are
interested in the thermodynamic limit $m \to \infty$. Using expression
(\ref{gcpfb}) for the partition function, we get:
\begin{equation}
\varphi_b=-\frac{1}{2}\ln \left\{\left(\frac{g_0^\prime}{g_0^{q-1}}
\right)^q\frac{1}{\left[1+qz_eR+\frac{q(q-1)}{2}z_iR^2\right]^{q-2}}
\right\},
\end{equation}
and recalling the recursion relation for $g_0$ (expression (11)),
which is:
\begin{equation}
g_0^\prime=g_0^\sigma+\sigma z_eg_0^{\sigma-1}g_1+
\frac{\sigma(\sigma-1)}{2}z_ig_0^{\sigma-2}g_1^2,
\label{rrg0b}
\end{equation}
we reach the following expression for the bulk free energy per site:
\begin{equation}
\varphi_b=-\frac{1}{2}\ln \left[ \frac{\left( 1+\sigma z_eR+
\frac{\sigma(\sigma-1)}{2}z_iR^2\right)^q}{\left( 1+qz_eR+
\frac{q(q-1)}{2}z_iR^2\right)^{q-2}}\right].
\end{equation}
Now the entropy is given as a state equation associated to the free
energy, whose expression is equation (\ref{ent}) above. since the
activities may be written as functions of the densities in the core of
the tree, expression (25) in \cite{ns08} for the entropy as a function
of these densities is found.

\end{document}